\documentclass[showpacs,preprintnumbers,pre]{revtex4}
\usepackage{graphicx}
\usepackage{dcolumn}
\usepackage{latexsym,amsmath,amssymb,graphics,color}
\usepackage{bm}
\usepackage[centerlast,footnotesize,sf]{caption}
\begin{document}

\title{Monte Carlo Studies of the Ising Antiferromagnet with a\\
Ferromagnetic Mean-field Term}

\author{Gregory Brown}
\affiliation{Computational Science and Mathematics Division,
Oak Ridge National Laboratory, Oak Ridge, Tennessee 37830, USA}

\author{Per Arne Rikvold}
\affiliation{Department of Physics, Florida State University,
Tallahassee, Florida 32306, USA}

\author{Seiji Miyashita}
\affiliation{Department of Physics, Graduate School of Science,
The University of Tokyo, 7-3-1 Hongo, Bunkyo-Ku, Tokyo, 113-8656, Japan} 

\begin{abstract}
The unusual thermodynamic properties of the Ising antiferromagnet supplemented with a
ferromagnetic, mean-field term are outlined. This simple model is inspired by more
realistic models of spin-crossover materials. The phase diagram is estimated using
Metropolis Monte Carlo methods, and differences with preliminary Wang-Landau Monte
Carlo results for small systems are noted.
\end{abstract}




\maketitle


\section{Model and Motivation}

Physical problems described by Hamiltonians composed of competing terms can display
very complicated behavior, and are thus quite interesting. Competition between
short-range and long-range interactions are particularly interesting, and have been studied
in the context of elastic long-range interactions which favor ferromagnetic (FM) ordering. If
the short-range interaction is also ferromagnetic, the critical behavior of the physical system
belongs to the mean-field universality class 
(Miyashita 2008, Nakada 2001).
If the short-range
order is antiferromagnetic (AFM) the statistical mechanics of the critical behavior is much more complicated.
This competition is essential to the molecular crystals known as spin-crossover materials 
(Miyashita 2008, Nakada 2011, Nishino 2013),
in which there is a volume change associated with the spin state of the molecules. 
Following Nakada {\em et al.} (2011), 
who studied the FM short-range interaction case, the case of an AFM
short-range interaction 
(Nishino 2013) 
competing with a long-range FM interaction in the 
Ising model is considered here. 

\begin{figure}[tb]
\begin{center}
\includegraphics[width=0.95\textwidth]{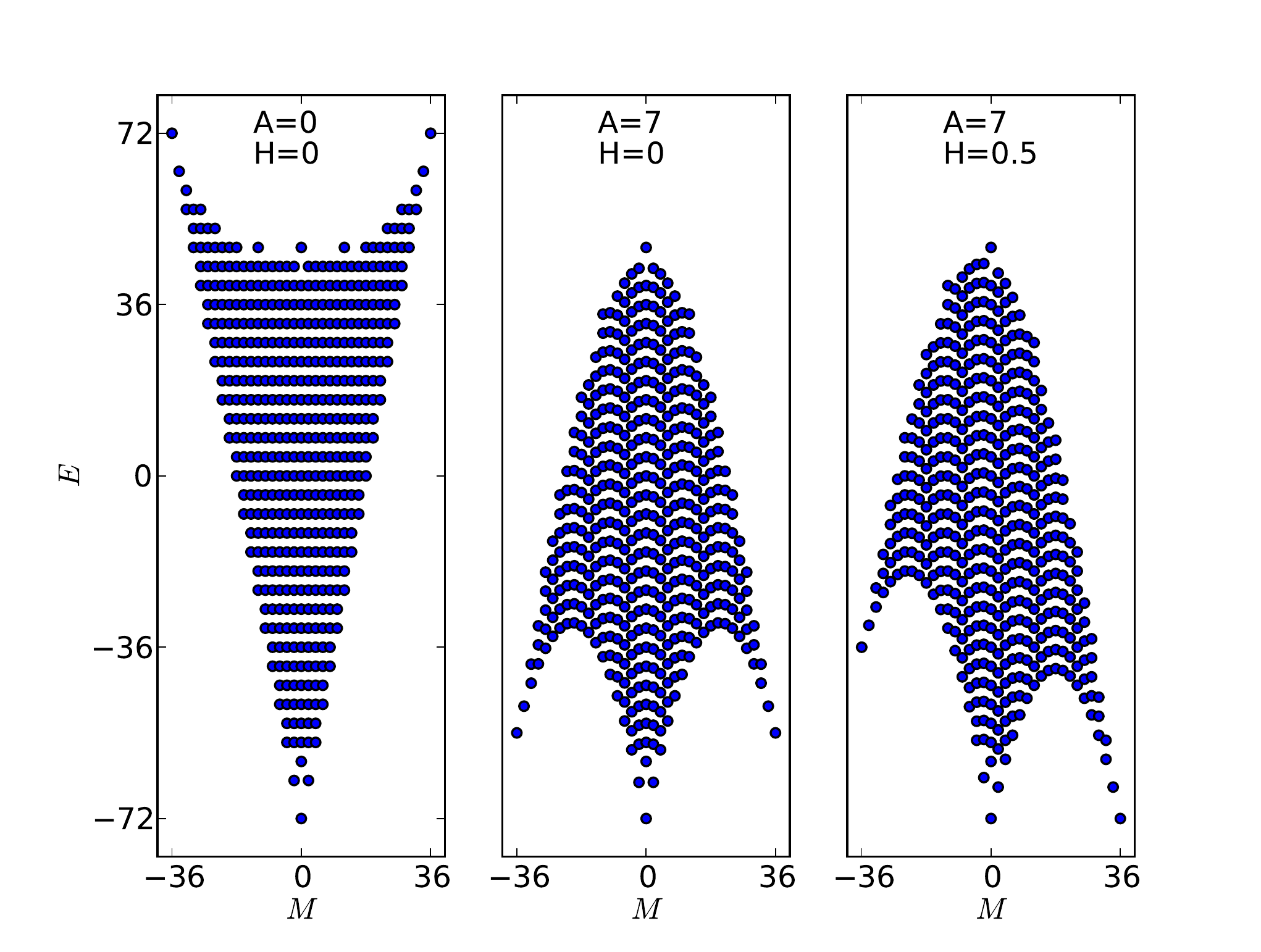}
\end{center}
\vskip  -0.05in
\caption[]{
Configuration classes of the $6\! \times\! 6$  Ising antiferromagnet for different
values of  mean-field coupling, $A$, and applied field, $H$. The left-most panel is the pure 
Ising antiferromagnet.
The middle panel shows the 
arrangement for $A\!=\!7$ ($H\!=\!0$), which brings the ferromagnetic
configurations to much lower energies.  The right-most panel shows the asymmetry introduced by
an external field $H\!=\!0.5$ ($A\!=\!7$).}
\label{fig:config}
\end{figure}

The model Hamiltonian is defined for Ising spins $s_i \! \in \! \{-1,+1\}$ 
arranged on a square $L \! \times \! L = N$ lattice
\begin{equation}
{\cal H} = -J \sum_{\langle i,j \rangle} s_i s_j - \frac{A}{2N} M^2 - H M
\label{eq:ham}
\end{equation}
with the sum restricted to nearest-neighbor pairs and $M\!=\!\sum_i s_i$ the magnetization
of the system. The second term is a long-range interaction of the mean-field 
Husimi-Temperley form 
(Miyashita 2008) 
and  favors FM configurations over AFM ones.
The third term is the normal Zeeman energy and  favors $s_i\!=\!+1$ for $H\!>\!0$. 
To study the antiferromagnet, it is sufficient to 
consider $J\!=\!-1$ with $A$, $H$, and the temperature $T$ in units of $|J|$. 

A configuration class is the subset of Ising configurations that have identical
energy $E$ and magnetization $M$ 
(Lourenco 2012). 
The configuration classes
for the $6 \! \times \! 6$ instance of Eq.~(\ref{eq:ham}) are shown in Fig.~\ref{fig:config}
for different values of $A$ and $H$. Changing the values of these parameters does not
change the membership of the  classes, but does affect the position of the  classes in the $E$-$M$ plane. The pure antiferromagnet, $A \! = \! 0$,
is shown in the left-most panel, while
the pure ferromagnetic Ising model is the reflection about the $x$-axis. 
At the critical field $H_c \! = \! 4$, the right-edge
of classes spanning the AFM configurations and the FM+ configuration are
all degenerate at $E \! = \! -2N$ for the former. Such degeneracy never arises
for the latter model, and this is one reason the AFM Ising model
has more complicated behavior 
(Hwang 2007).

The effect of increasing $A$ is much different. The middle panel of Fig.~\ref{fig:config}
shows the arrangement of configuration classes for $A \! = \! 7$, which is the value considered
throughout this paper. The arrangement looks similar to the FM Ising model, with curved
boundaries separating the classes locally minimizing $E$. For $A \! = \! 8$ the AFM and FM 
classes are exactly degenerate. The right-most panel shows the effect of applying a field
on the arrangement of the classes. Clearly, this shows that depending on the values of
$A$ and $H$ the phase transition between the AFM and paramagnetic (PM) states can be either first order
or second order.

\begin{figure}[tb]
\begin{center}
\includegraphics[width=0.95\textwidth]{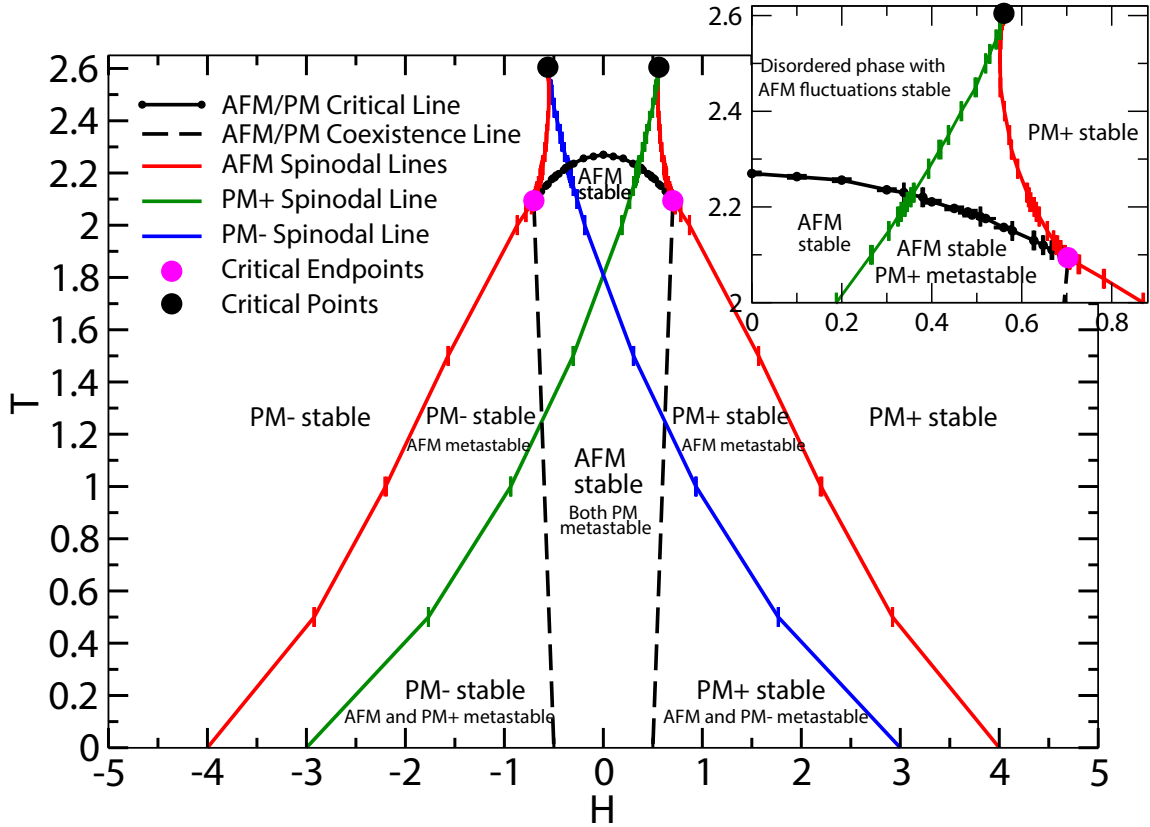}
\end{center}
\vskip  -0.05in
\caption[]{
Phase diagram for the mean-field Ising antiferromagnet with $A\!=\!7$ as determined by Metropolis
Monte Carlo. The black curve with small circles at data points represents critical temperatures determined by extrapolation of cumulant crossings. 
The vertical ticks are spinodal points estimated from Monte Carlo scans in $H$; the 
horizontal size of the ticks indicate the uncertainty in the estimate. The points are connected with
straight segments as a guide to the eye.}
\label{fig:phase}
\end{figure}

\section{Results}

The phase diagram for $A\!=\!7$ deduced from Metropolis Monte Carlo 
(Metropolis 1953)
simulation scans in $T$ and $H$
is shown in Fig.~\ref{fig:phase}, along with the relevant spinodal lines. At $T\!=\!0$, first-order 
transitions between AFM and FM order occur at $|H|\!=\!0.5$. The distinction between FM configurations
and PM configurations is rather arbitrary because there is not always a phase 
transition separating them.
  For $T\!>\!0$, coexistence curves for AFM/FM order
are sketched approximately as dashed black lines. 
The curve of second-order AFM/PM phase transitions measured via the Binder
cumulant method is shown as points connected by solid line segments.
These extend to $H\!>\!0.5$, indicating slight reentrance of the AFM phase. 
Vertical ticks represent
numerical determination of the spinodal lines 
(Miyashita 2008) 
including error estimates,
with line segments included as guides to the eye.
Below all the curves the AFM configurations are minima in the free energy: global minima and 
absolutely stable within the curves of the phase transitions, as opposed to metastable minima 
between the transition lines and the spinodals. The positively oriented PM$+$ configuration has a free-energy
minimum to the right of the PM$+$ spinodal (green), while the negatively oriented PM$-$ configuration
has a free energy minimum to the left of the PM$-$ spinodal line (blue). 

Aside from the overall temperature scale, the features of the phase diagram considered so far are 
similar to those for a model in which the local AFM interaction term is replaced by the
two-sublattice, mean-field approximation, {\em i.e.} the first term of Eq.~(\ref{eq:ham}) is
replaced by $2|J|m_{\rm A} m_{\rm B}$ with $m_A$ and $m_B$ the individual sublattice
magnetizations. In the mean-field approximation,  the coexistence line joins the line of critical 
points at a tricritical point,  from which the AFM and PM spinodal lines also emanate. 
However, the local nature of the AFM term in 
Eq.~(\ref{eq:ham}) admits fluctuations that significantly change the topology of the phase 
diagram at higher temperatures. Instead of merging at a tricritical 
point, the coexistence and second-order lines meet at an angle
in a {\em critical endpoint} 
(Fisher 1990, Tsai 2007)
at approximately $H\!=\!0.70$ and $T\!=\!2.09$,
shown  in the inset in Fig.~\ref{fig:phase}. 
Above the temperature of the critical endpoint, the coexistence line 
continues to a critical point at approximately $H\!=\!0.561$ and $T\!=\!2.605$. 
The FM+ spinodal line 
also ends at this point. In the application of this model to spin-crossover materials,
such a phase diagram provides for a variety of interesting hysteresis behaviors. 

\begin{figure}[tb]
\begin{center}
\includegraphics[width=0.95\textwidth]{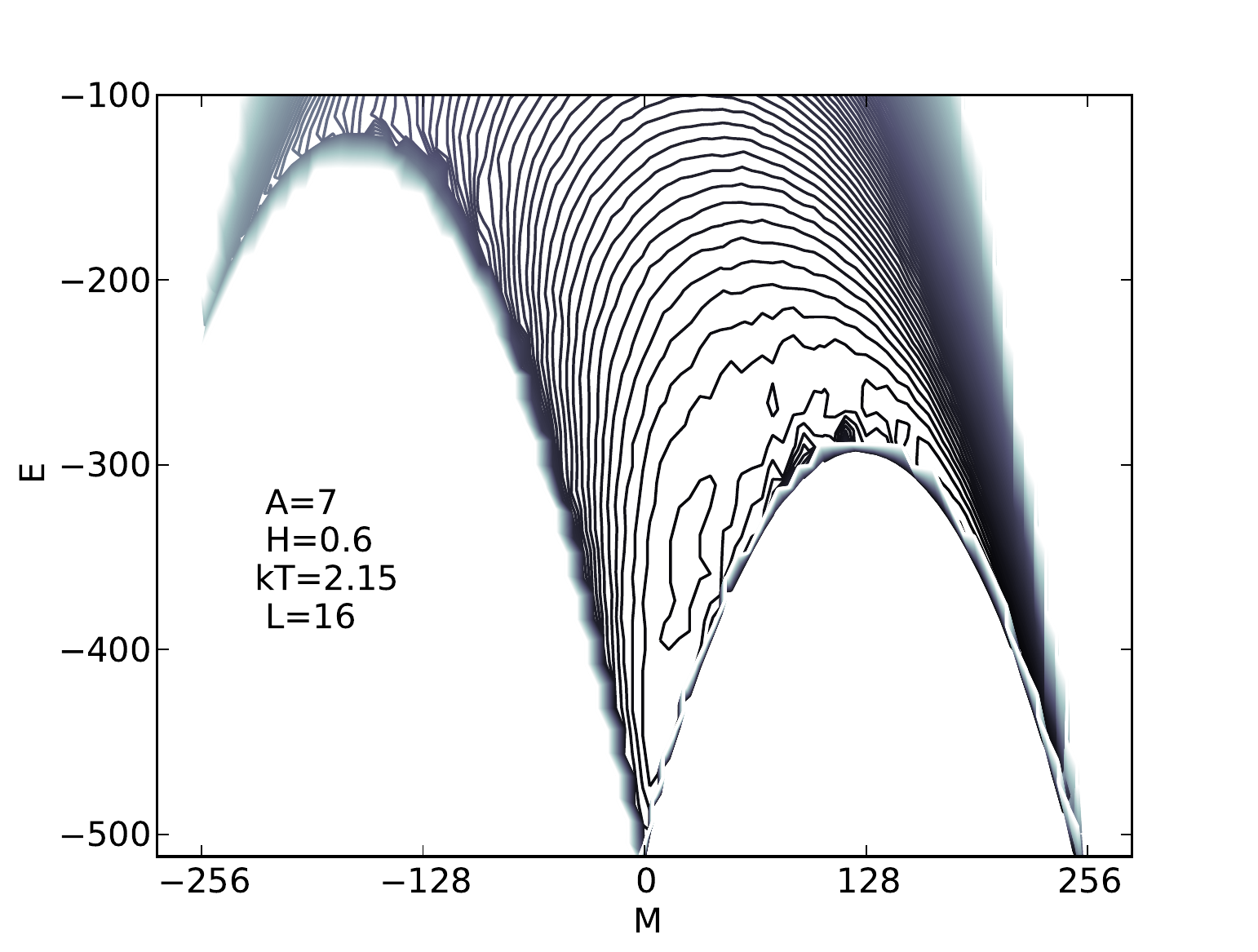}
\end{center}
\vskip  -0.05in
\caption[]{
The free energy $E-TS$ calculated from the ``two-dimensional'' density of states, $g(E,M)$
for $A\!=\!7$ and $H\!=\!0.6$ for $T\!=\!2.15$, near the critical endpoint in the inset of Fig~2.
The stable minimum is in the AFM region, but it is shallow and extends nearly to the PM+ region.}
\label{fig:freecont}
\end{figure}

To understand the origin of these details, the Wang-Landau Monte Carlo method 
(Wang 2001)
has been
employed to estimate the density of states, $g(E,M)$. The entropy of the Hamiltonian is
$S(E,M) \! = \! \ln{[g(E,M)]}$ and so the free energy is
$
F(E,M) = E - T \ln{\left[ g \left(E,M \right) \right ]}
$ 
since all energies, including $T$, are in units of $|J|$. A contour plot of the free energy
is shown in Fig.~\ref{fig:freecont} for $L\!=\!16$, $A\!=\!7$, $H\!=\!0.6$, and $T\!=\!2.15$.
These conditions are near the critical endpoint for this small system, but with the global minimum clearly 
associated with AFM configurations. As $H$ is increased this shallow minimum passes over to
the region associated with FM/PM configurations. However, the phase diagram determined by
the maximum in the specific heat from the Wang-Landau data (for this very small $L$)
does not agree with the Metropolis data in several respects.
The coexistence line of first order AFM/FM phase transitions
rises vertically at $H\!=\!0.5$, before curving to the left for temperatures significantly above
$T\!=\!1$ for $L\!\le\!16$. No reentrant behavior has been observed in the Wang-Landau
data. Instead, the maximum in the specific heat is observed to bifurcate at a field less than
$H\!=\!0.5$. The left-most maximum corresponds the line of AFM/PM phase transitions, while the right-most maximum does not scale with the system size and possibly corresponds to a disorder line in the PM phase.

\section{Summary}

Competition in the antiferromagnetic Ising model supplemented with a mean-field ferromagnetic term,
leads to very rich thermodynamic behavior. One interesting feature is the transformation of a tricritical point
(in the purely mean-field model) into a critical endpoint (in the full model). Preliminary results for Wang-Landau Monte Carlo on very small systems have not yet been harmonized with extensive results from Metropolis Monte Carlo calculations, particularly near the critical endpoint. 

\vskip  0.05in
GB is supported by Oak Ridge National Laboratory, which is managed by UT-Battelle, LLC.
PAR acknowledges hospitality at The University of Tokyo and partial support by NSF Grant No. DMR-1104829.

\begin{small}



\vskip  0.05in

\noindent
Fisher, M.E., Upton, P.J., 1990, Phys.\ Rev. Lett.\ {65}, 2402. 

\noindent
Hwang, C.-O., Kim, S.-Y., Kang, D., Kim, J.M., 2007,
J.\ Stat.\ Mech.:\ Theory and Experiment {2007}, L05001.

\noindent
Lourenco, B.J., Dickman, R., 2012,
Int.\ J.\ Mod.\ Phys.\ C {23}, 1240007.

\noindent
Metropolis, N., Rosenbluth, A.W., Rosenbluth, M.N.,  Teller, A.H., Teller, E., 1953,
J.\ Chem.\ Phys.\ {21}, 1087.

\noindent
Miyashita, S., Konishi, Y., Nishino, M., Tokoro, H., Rikvold, P.A., 2008,
Phys.\ Rev.\ B {77}, 014105. 

\noindent
Nakada, T., Rikvold, P.A., Mori, T., Nishino, M., Miyashita, S., 2011,
Phys.\ Rev.\ B {84}, 054433.

\noindent
Nishino, M., Miyashita, S., 2013,
Phys.\ Rev.\ B {88}, 014108.

\noindent
Tsai, S.-H., Wang, F., Landau, D.P., 2007, Phys.\ Rev.\ E {75}, 061108. 

\noindent
Wang, F., Landau, D.P., 2001,
Phys.\ Rev.\ Lett.\ {86}, 2050; Phys.\ Rev.\ E {64}, 056101.

\end{small}

\end{document}